\documentclass[12pt,a4paper]{article}
\usepackage[latin2]{inputenc}
\usepackage{amssymb,amsmath}
\usepackage{graphicx,float}

\begin{document}
\title{Preferential Behaviour and Scaling in Diffusive Dynamics on  Networks}
\author{Bernard Kujawski$^{1}$, Bosiljka Tadi\'{c}$^{2}$ and G. J. Rodgers$^{1}$\\
$^{1}$Department of Mathematical Sciences, Brunel University, Uxbridge;\\
Middlesex UB8 3PH; UK \and
$^{2}$Department for Theoretical Physics, Jo\v{z}ef Stefan Institute, \\
P.O. Box 3000, SI-1001 Ljubljana; Slovenia }

\maketitle
\begin{abstract}
We study the fluctuation properties and return-time statistics on
inhomogeneous scale-free networks using packets moving with two
different dynamical rules; random diffusion and locally navigated
diffusive motion with preferred edges. Scaling in
the fluctuations
occurs when the dispersion of a quantity at each node or edge
increases like the its mean to
the power $\mu$. We show that the occurrence of scaling in the
fluctuations of both the number of packets passing nodes and the
number flowing along edges is related to preferential behaviour in
either the topology (in the case of nodes) or in the dynamics (in
case the of edges). Within our model the absence of any preference
leads to the absence of scaling, and when scaling occurs it is
non-universal; for random diffusion the number of packets passing
a node scales with an exponent $\mu$ which increases continuously
with increased acquisition time window from $\mu =1/2$ at small
windows, to $\mu =1$ at long time windows; In the preferentially
navigated diffusive motion, busy nodes and edges have exponent
$\mu =1$, in contrast to less busy parts of the network, where an
exponent $\mu =1/2$ is found. Broad distributions of the return
times at nodes and edges illustrate
that the basis of the observed scaling is the cooperative behaviour between groups of nodes or edges. These conclusions are relevant
for a large class of diffusive dynamics on networks, including
packet transport with local navigation rules.
\end{abstract}

\section{Introduction}
In transport processes on networks complex dynamical behaviour
may  be caused by the structure of underlying network
geometry (for a recent review see \cite{trt-tcnfjo-06} and
references therein). The inhomogeneous degree distributions found in the scale-free networks may be one cause of this behaviour, however, other structural details appear to be essential
when dynamics include the local navigation of packets
\cite{trt-tcnfjo-06,tt-isdsn-04,ttr-tcntugspmdf-04}. Queuing effects at nodes, which is essential for high density information
packet transport
\cite{trt-tcnfjo-06,tt-isdsn-04,ttr-tcntugspmdf-04,gadg-dpmcn-02,gdvca-ontlsc-02,sv-itptmit-01}, and  is usually
absent in other transport processes such as charge transport
\cite{st-tphpgsfl-06}, is another cause of collective behaviour in transport dynamics.

Many dynamically measurable  outputs exhibit scaling features,
indicating a degree of universality that, in turn, can be used to
probe a network's structural and traffic properties
\cite{trt-tcnfjo-06}. Distributions of transport times, waiting
times and noise fluctuations exhibit power-law behaviour in
different network models. However, it is not an easy task to
relate the emergent dynamical features to particular structural
properties, firstly because of the network's power-law
inhomogeneity but also because different structural elements,
i.e., nodes or edges, or higher (hidden) structures, may play a
role in the dynamics. Therefore, a more systematic study of
traffic scaling properties,
 that considers structure beyond the node's connectivity, is necessary in
order to reveal the origin of scaling and relate the observed
scaling properties to the specific structural and dynamical
features on complex networks.

In view of the long-range correlations in packet streams on
inhomogeneous
 networks, analysis of  {\it traffic
noise}, defined by the number of packets processed by a node, and
{\it traffic flow}, the number of packets passing along an edge,
can  give interesting information about traffic conditions and the
underlying network structure \cite{trt-tcnfjo-06}.
 In recent studies a multi-channel analysis of traffic noise on networks
\cite{mb-fnd-04,da-otusftcn-06,t-sfnfsfn-06} reveals that the set of
fluctuations $\{\sigma _i\}$ of the traffic time-series
$\{h_i(t)\}$  measured at all nodes in the network $i=1,2, \cdots N$
obeys the scaling law
\begin{equation}
\sigma _i \sim <h_i>^{\mu} \ . \label{eq-sigmah}
\end{equation}
The exponent $\mu$ is often found to be either $\mu =1/2$ or $\mu
=1$, suggesting a super-universal behaviour across different
networks \cite{mb-fnd-04}. However, recently it was found
\cite{da-otusftcn-06,t-sfnfsfn-06,trt-tcnfjo-06} that the exponent
$\mu$ may depend on traffic conditions and the type of
measurements taken. Similar results, in which non-universal
scaling was found to depend on the acquisition time window,  where
found in the analysis of stock market time series
\cite{ek-sttcsdftvs-06}. A unique exponent between these two universal
values was also found in the analysis of the genome-wide time series
of the gene expression
 of yeast \cite{ztwt-sicbfcgeny-06}, the dynamics of which
is naturally limited by the cell cycle.

In this paper  we address the question of scaling in the diffusive
dynamics on networks, by carefully selecting the inhomogeneity of
the network and the dynamical rule, in order to determine the
origin of the scaling and its robustness.
To demonstrate the importance of
these dynamic phenomena in real-world networks, we employ a
non-trivial model for the transport of information packets
\cite{tr-ptsfn-02,tt-isdsn-04,ttr-tcntugspmdf-04,trt-tcnfjo-06}.
For the purpose of this work we use a simple scale-free network
\cite{ab-smcn-02,dm-en-03} with a clear inhomogeneity in the node connectivity
but with low clustering and no edge
correlations. We consider two types of diffusive motion of the
packets on the graph: random diffusion and local navigation with
preferred edges, as described below.

In comparison to the models considered in
\cite{mb-fnd-04,da-otusftcn-06}, our traffic model is more realistic
in that:
\begin{itemize}
    \item Packets are created at a given rate $R$ and travel to specified destinations on network;
\item Packets queue dynamically at nodes;
\item Packets are navigated locally according to a specified algorithm.
\end{itemize}
Consequently the travel-times are determined
self-consistently rather than being fixed as an adjustable
external parameter of the model and
have a broad distribution with a
power-law tail, which depends on the network structure
navigation algorithms \cite{tt-isdsn-04,tt-statsfn-05}.
The waiting times of packets in
queues is another property of our model that is determined
self-consistently by the dynamics. Depending on the type of
queuing discipline employed, the waiting time distribution can
also have a power-law tail on scale-free networks
\cite{tt-isdsn-04}. Throughout this work we use low packet
densities in order to keep the time series of traffic noise and
flow stationary and avoid the effect of large queuing times (see
\cite{trt-tcnfjo-06} for a study of dense traffic on structured
networks).

Our main findings suggest that the presence of a  preference in
either the topology, such as in node connectivity, or in the dynamics, such as
by edge-preferred navigation, leads to the scaling of
fluctuations. When this scaling occurs, careful analysis shows
that it is non-universal, depending both on the
acquisition-time and on the importance of the nodes or edges in
the transport process. These findings of collective dynamical behavior
on networks are further substantiated with
a study of the return-times statistics for nodes and edges.

The organization of this paper is as follows. In section 2 we
define the network and the traffic model and summarize the main
features of the transport process. In section 3 the results for
noise and flow fluctuations are given for various parameters and
dynamic rules and in section 4 the origins of scaling are
discussed and compared with  transport on much simpler geometries.
In section 5 the results for the return-times distributions are
obtained for the network dynamics studied in section 3, and
section 6 gives a short summary and discussion of our results.

\section{Traffic of information packets with queuing and navigation on network}

\subsection{\it Network structure}
We consider a simple scale-free network \cite{ab-smcn-02,dm-en-03} with a
power-law inhomogeneity in node connectivity, low clustering and
no correlations between edges. Such a network can be easily grown with a
preferential attachment rule in which a node with two links is
added to the network at each time step and each link is connected
to a node of degree $k$ with a rate proportional to $k+\alpha$.
In the emergent structure the degree of the
$i^{th}$ added node $k_i \sim (i/N)^{-1/(1+\alpha)}$, and hence the
connectivity distribution has a power-law tail, $P(k) \sim
k^{-\tau}$ with $\tau = 2 + \alpha$, as shown analytically \cite{dm-en-03}.
 A detailed analysis of the structure of the networks grown in this way
shows that
the clustering is very low and link correlations are entirely
absent, in contrast to correlated scale-free networks, grown with the
algorithms  described in \cite{t-arwcwg-01}.
For our simulations of packet diffusion we grow a simple uncorrelated
scale-free
network of $N=1002$ nodes and $E=2N$ edges, which has a connectivity
distribution with a power-law exponent $\tau \approx 2.5$.

Once the network is generated, we consider it's structure as  fixed,
and start the  transport processes on it. During these dynamics, at each
time step  each node can
create a new packet with
probability $R/N$. At creation each packet is assigned a
destination address, another node on the network where it should
be delivered.  Packets move diffusively through the network, either
performing a random walk or navigated
according to the algorithm described below.
Once at the destination address packets disappear
from the network.  In this model packets move towards their
destination simultaneously, forming queues at the nodes on the way. We
assume a finite queue buffer of length $H=1000$ at each node. The
first in - first out (FIFO) queuing discipline is applied. The
maximum length of the queue is important for transport close to the
congestion state and has an impact on the scaling of the
fluctuations on nodes \cite{trt-tcnfjo-06,t-sfnfsfn-06}.

\subsection{\it Random diffusion and edge-preferred navigation}

The motion of packets through the network can either be random or
navigated using some rules, which may affect the role of different
nodes and edges in the transport process. In the case of random
diffusion (RD) a node that is processing a packet selects
one of its neighbouring nodes at random to send the packet to it.
Hence the nodes on the network
are visited at a rate proportional of their connectivity, and hence
nodes with a high connectivity are often visited by moving packets.
Implementation of a navigation
algorithm \cite{ktr-libaptcn-06,trt-tcnfjo-06} will create a bias
in the use of
nodes and edges on a given topology. Here we consider one such
navigation algorithm, referred to as the CD algorithm,
 that introduces a preference for less used
edges in the traffic history \cite{ktr-libaptcn-06}. In particular,
a node $i$ that is
processing a packet at time step $t$ selects one of its neighbour
nodes, $j$, that has the {\it minimum }value of the quantity
$s_{ij}(t)$ defined as the product
\begin{equation}
s_{ij}(t)=k_jF_{ij}(t) \ , \label{CD-def}
\end{equation}
where $k_j$ is the degree of the node  $j$ and
$F_{ij}(t)$ is the number of packets forwarded from node $i\to j$
up to the time $t$. This means that the CD algorithm is completely
deterministic, involves a search of the nearest neighbourhood
of the processing node, prefers to send packet to nodes away from
the hubs and hence introduces a dynamical inhomogeneity to the network
transport. In this way,
from the point of view of nodes, the topological inhomogeneity of
the network appears to be dynamically reduced. At the same time,
an unequal use of edges appears, that is measured by the packet
flow on them. Note that this property of the navigation rules may
reduce jamming problems at high traffic density, which often occur
on scale-free networks in algorithms based on the shortest
paths \cite{ktr-libaptcn-06}.
We consider low packet density by keeping
a low posting rate $R$, which is much below the jamming rate for
this network's structure and navigation rules, in order to minimize
the potential effects of
long queues and to retain the stationarity of the traffic time
series.

\subsection{\it Traffic properties}

\begin{figure}[htb]
\begin{center}
\begin{tabular}{cc}\hskip -0.5 cm
\includegraphics[width=7.4cm]{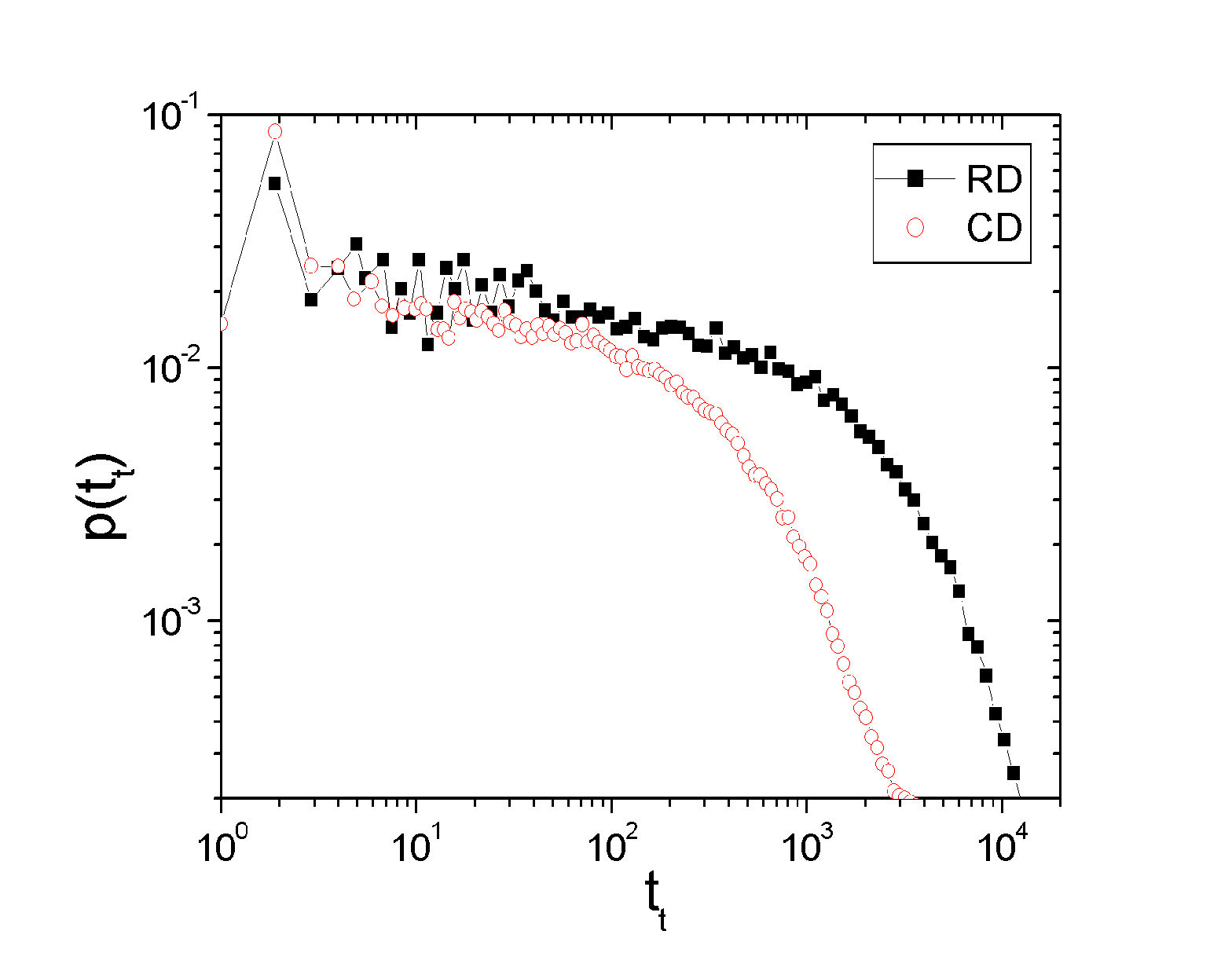} &
\includegraphics[width=6.6cm]{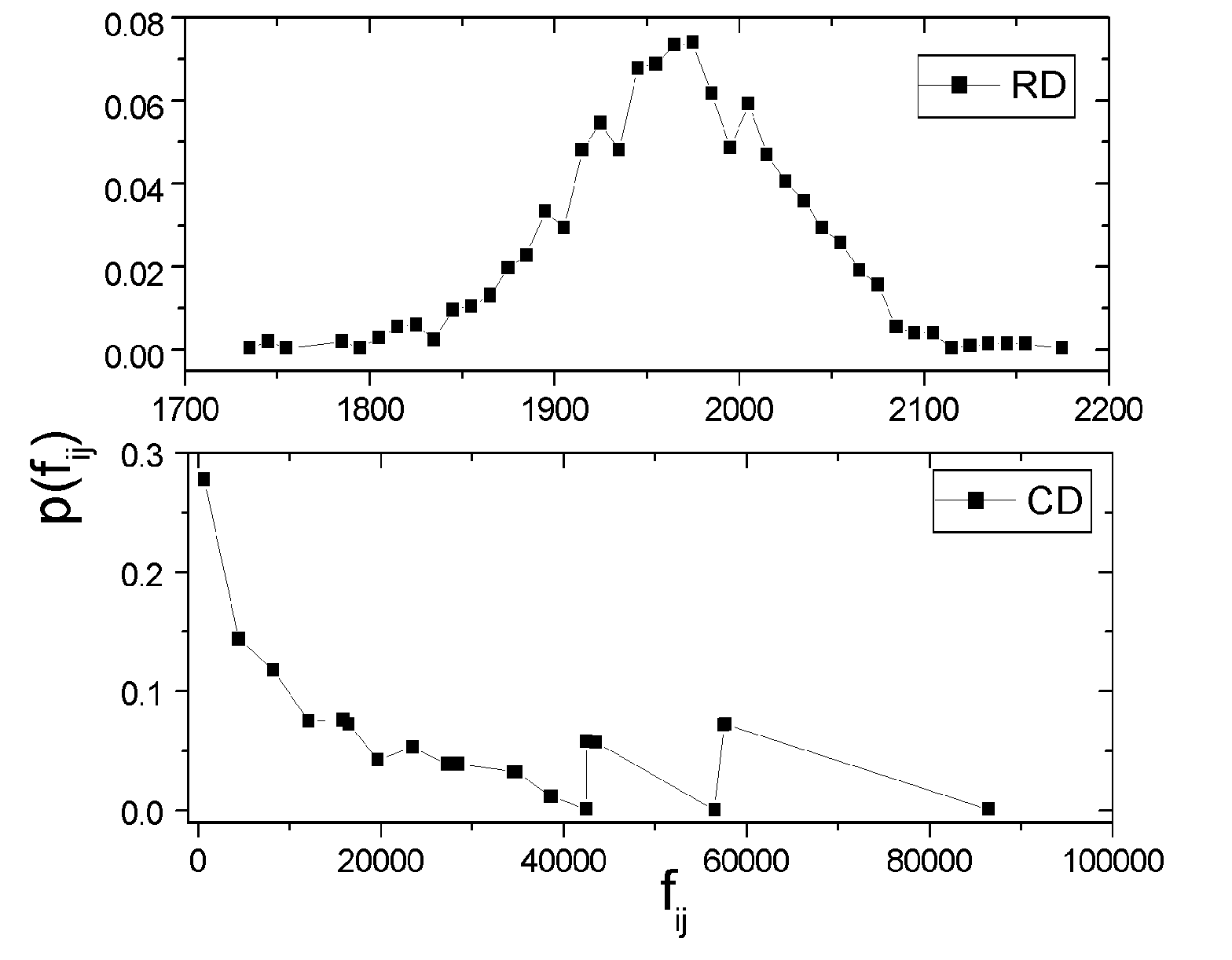}  \\
{$(a)$} &{$(b)$}  \\
\end{tabular}
\end{center}
\caption{(a) Distribution of travel times of packets and (b)
distribution of flow along the edges for both random diffusion (RD) and
edge-preferred navigation (CD).} \label{fig-traffic-properties}
\end{figure}

For the fixed network topology described above and a fixed posting
rate of packets $R$, we simulate packet transport both with random
diffusion rules and with the CD navigation algorithm.
The transport properties
are measured by a number of global and local statistical
quantities \cite{tr-ptsfn-02,tt-isdsn-04,ttr-tcntugspmdf-04,trt-tcnfjo-06},
which depend on the diffusion rules. Here we summarize several
traffic properties which are relevant for further discussion.
 In Fig.\ \ref{fig-traffic-properties} we show how the two diffusion rules effect the distributions of travel times of packets and flow along the edges on the scale-free network.
Both in the random diffusion and in the navigated transport the
distribution of travel times exhibits a power-law tail, however,
both slopes and cut-offs are different.
The distribution of dynamic flow along edges of the network is
shown in Fig.\ \ref{fig-traffic-properties}b. For RD the dynamical
flow is similar to the topological flow (the centrality measure of
edges) and is dominated by an average value with a width which depends
on the packet density. In contrast, the CD navigation algorithm induces a
non-symmetric flow distribution created by the dynamically preferred edges.

\begin{figure}[htb]
\begin{center}
\begin{tabular}{cc}
\includegraphics[width=6.8cm]{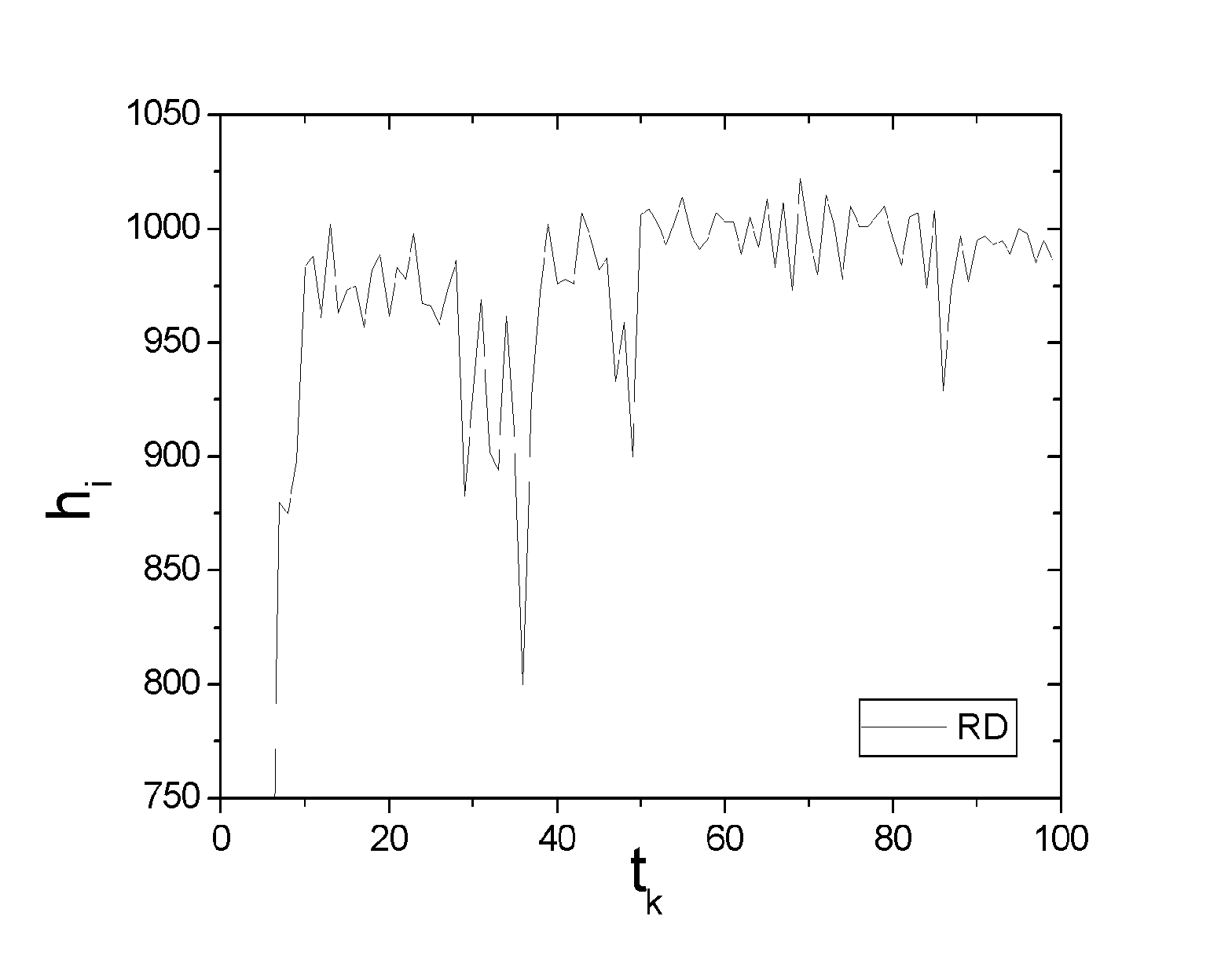}&
\includegraphics[width=6.8cm]{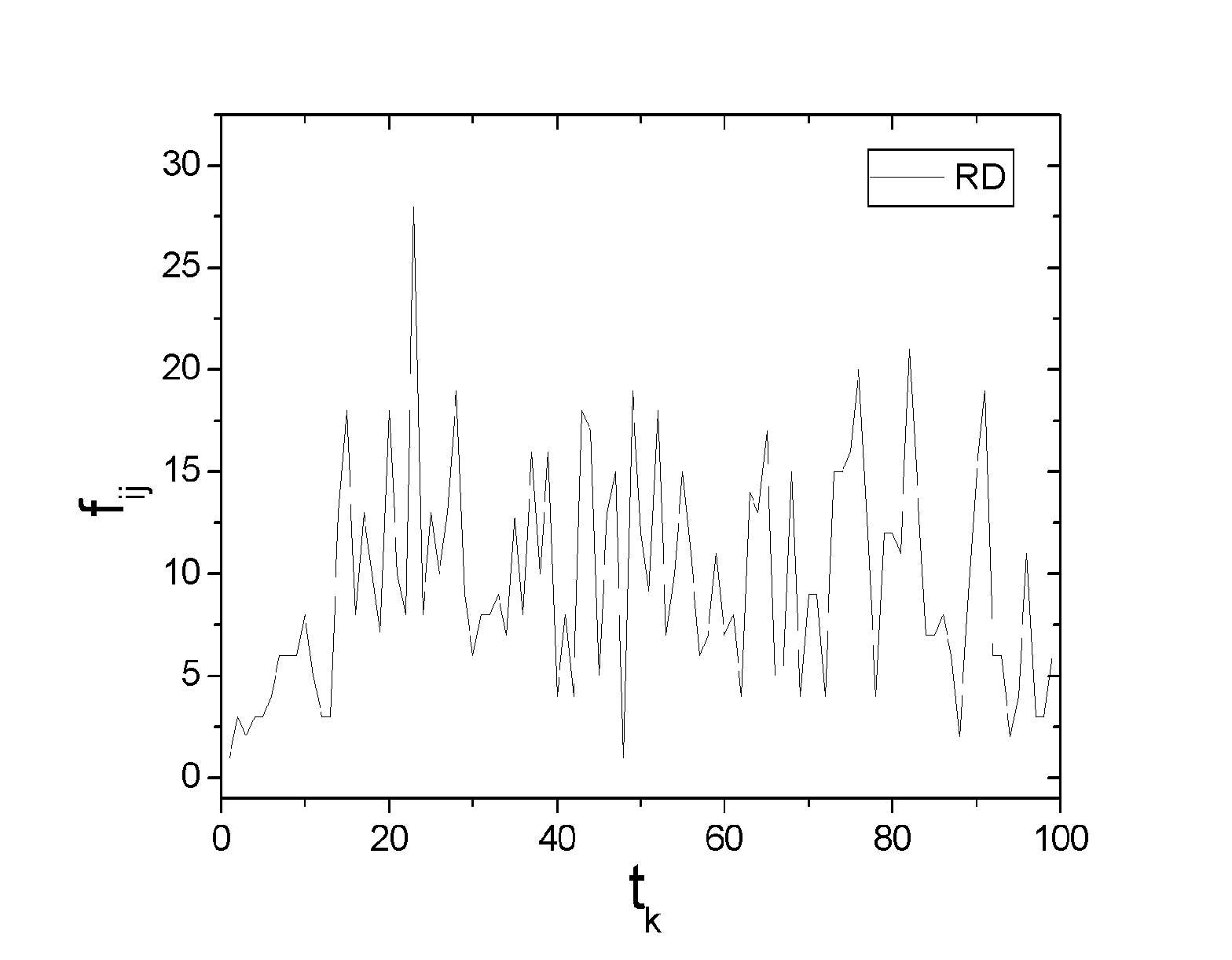}\\
{$(a)$} &{$(b)$}\\
\end{tabular}
\end{center}
\caption{Example of time series recorded at (a) a preferred node and (b) at an edge with random diffusion rules and  time-window $T_{WIN}=1000$ steps.
}
\label{fig-2timeseries}
\end{figure}

Further differences in traffic with the two diffusion rules are
shown in the time series of the number of packets processed
by individual nodes and edges, shown in Fig.\
\ref{fig-2timeseries}. A detailed analysis of such  time series is
given in the following section.

\section{Scaling of Noise and Flow}

Noise and flow are two local characteristics of the traffic that
are determined by the number of packets transported in a given time window $t_k$ at a node $i$
$h_i(t_k)$, and along a  link $i\to
j$ and $j\to i$, $f_{ij}(t_k)$. The index
$k=1, 2 \cdots K_{max}$ enumerates time windows of length
$T_{WIN}$ time steps.

The flow $f_{ij}(t_k)$ can be written as
\begin{equation}
f_{ij}(t_k)=F_{ij}(t_k)+F_{ji}(t_k)-[F_{ij}(t_{k-1})+F_{ji}(t_{k-1})]
\end{equation}
where $t_k=kT_{WIN}$. During the transport process we record a set of fluctuating
time-series, one for each of $N$ nodes $\{h_i(t_k)\}$, $i=1,2,
\cdots N$, and similarly another set of time-series collected for each
of the $2N$ edges $\{f_{ij}(t_k)\}$ on the network. In the simulations
we fix the creation rates of packets at $R=0.01$ for the RD and
$R=0.1$ for CD-navigated diffusion, which are well below the respective
jamming rates on this network structure.

\subsection{\it  Noise fluctuations at nodes with scale-free connectivity}

\begin{figure}[htb]
\begin{center}
\begin{tabular}{cc}
\includegraphics[width=6.8cm]{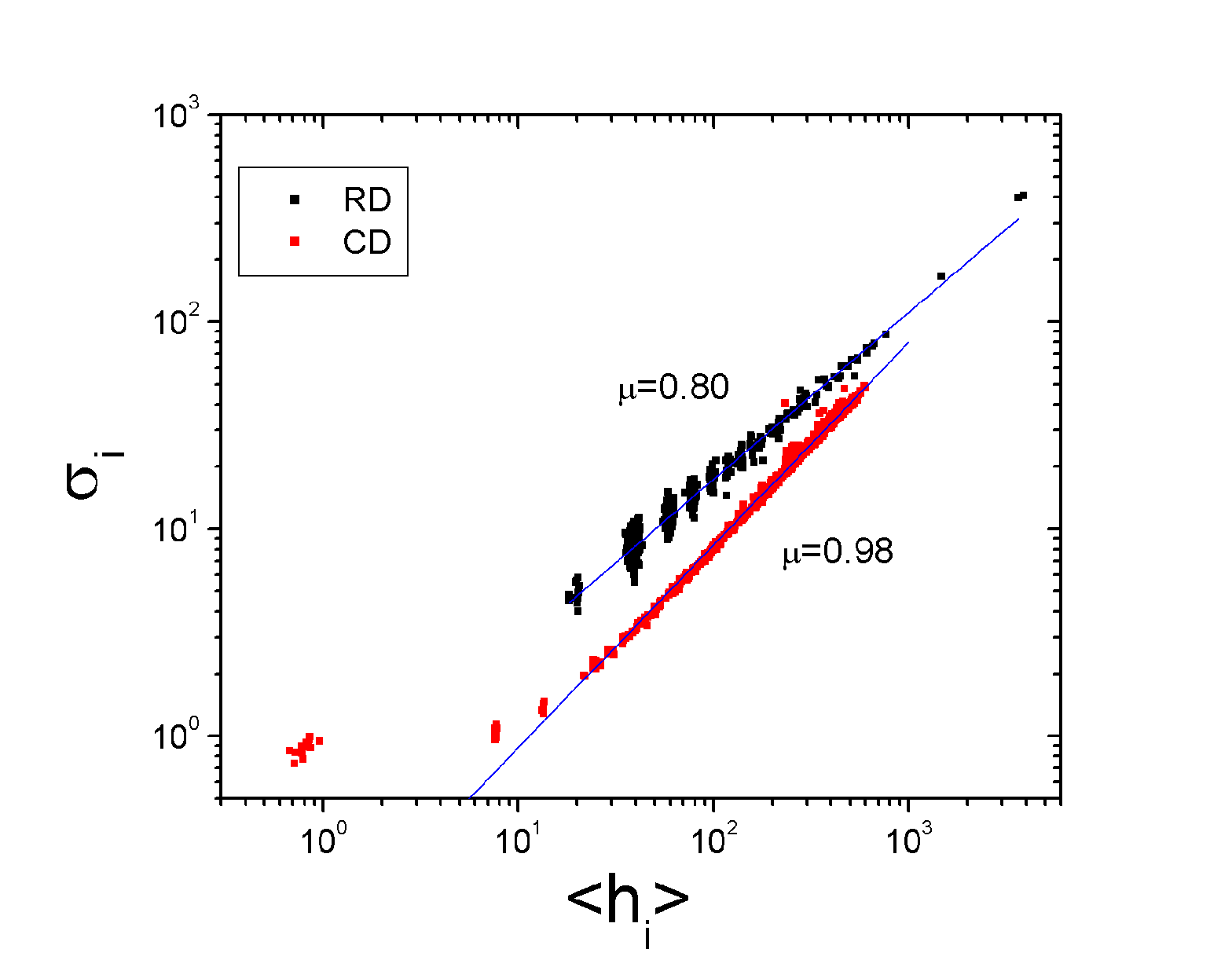}&
\includegraphics[width=6.8cm]{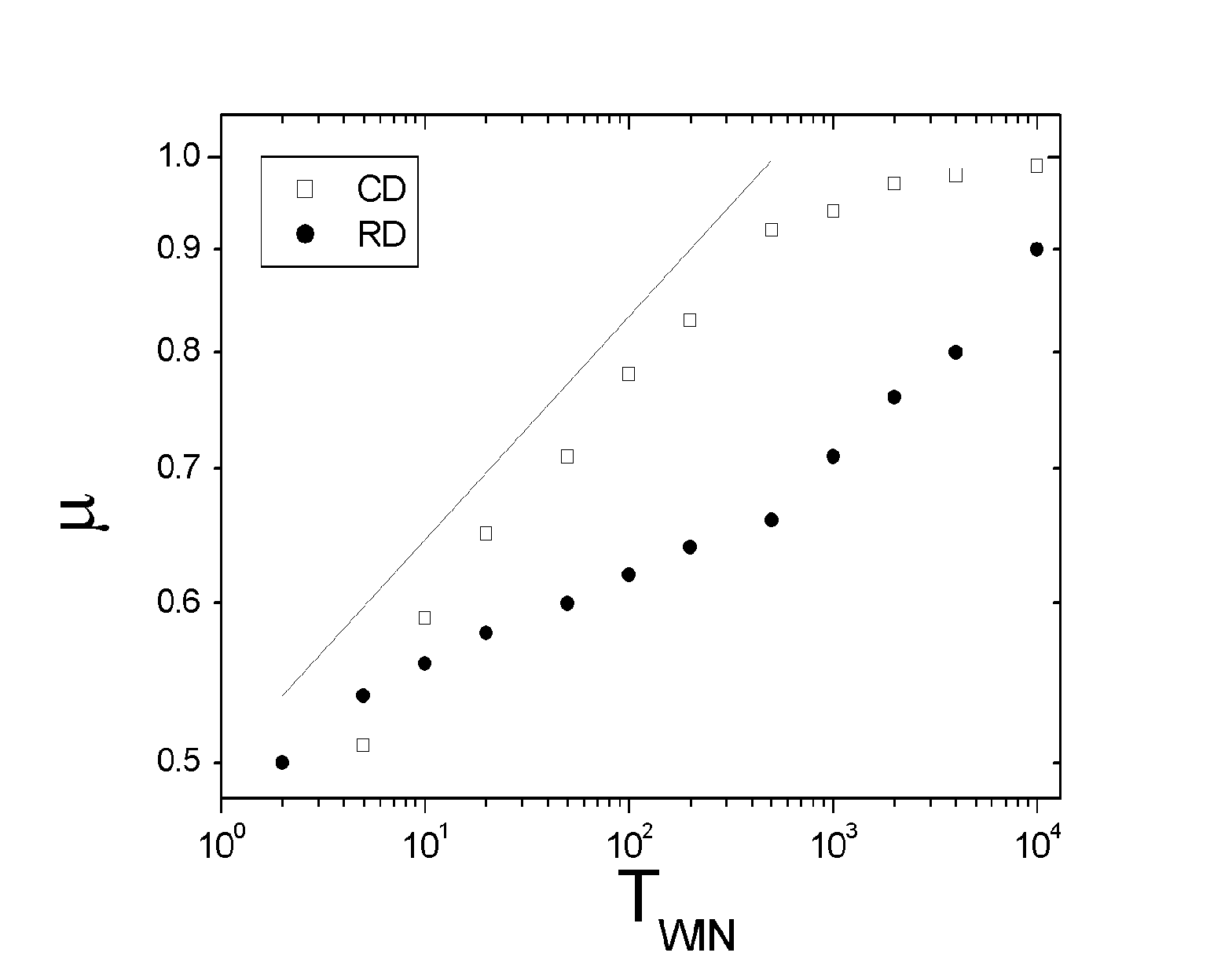}\\
{$(a)$} &{$(b)$}\\
\end{tabular}
\end{center}
\caption{ (a) Dispersion $\sigma _i$ against average $<h_i>$ of time series recorded at
nodes of the network within a fixed time window $T_{WIN}$ for random (RD) and
navigated (CD) diffusion. (b) Dependence of the scaling
exponent $\mu$ on the width of the time window.  Slope of the line is
$\beta=0.11$.} \label{fig-sigh-nodes}
\end{figure}

We determine the dispersion $\sigma _i$ and an average $<h_i>$ of
long time series recorded at all nodes  $\{h_i(t_k)\}$, $i=1,2,
\cdots N$ and $k=1,2, \cdots K_{max}$. Plots of $\sigma _i$
against $<h_i>$ are given in Fig.\ \ref{fig-sigh-nodes}a, where
each point represents one node of our scale-free network. The two
curves correspond to the results for RD dynamics
and for the transport with the CD navigation algorithm and a fixed
acquisition time window $T_{WIN}=4000$ time steps. As the Fig.\
\ref{fig-sigh-nodes}a shows, these plots obey the scaling equation
(\ref{eq-sigmah}) with a well defined exponent $\mu$, which
is different for the two algorithms.
For RD the value of $\mu$ is clearly between the two limits 1/2 and
1 mentioned above, whereas in the case of the CD navigation algorithm
 $\mu$ appears to be  close to 1.

By changing the width of the time window in which the data is
collected we find that the scaling law still holds, but with
a different exponent $\mu$.  This applies for both RD and CD, however,
the functional
dependencies $\mu(T_{WIN})$ are different. The results are
 shown in Fig.\ \ref{fig-sigh-nodes}b.
As a rule, if data are acquired in a longer time window, the
exponent $\mu$ tends to be larger.
For RD we
find the change from weak dependence at short time windows to
steeper increase when large windows are applied. For CD the exponent is
constantly large
for large time windows, and changes rapidly when the windows are
smaller than the typical $T_{WIN} \approx 500$ time steps in these
simulations. In fact, for CD we find that the
dispersion between the groups of nodes exhibiting the scaling with
large $\mu$ exponent is gradually reduced with decreasing time
window and eventually the group becomes so condensed that an
exponent cannot be defined. Simultaneously another group emerges
with the scaling exponent close to $\mu =1/2$.

\subsection{\it  Scaling of flow at preferred edges}

\begin{figure}[htb]
\begin{center}
\begin{tabular}{cc}
\includegraphics[width=7.0cm]{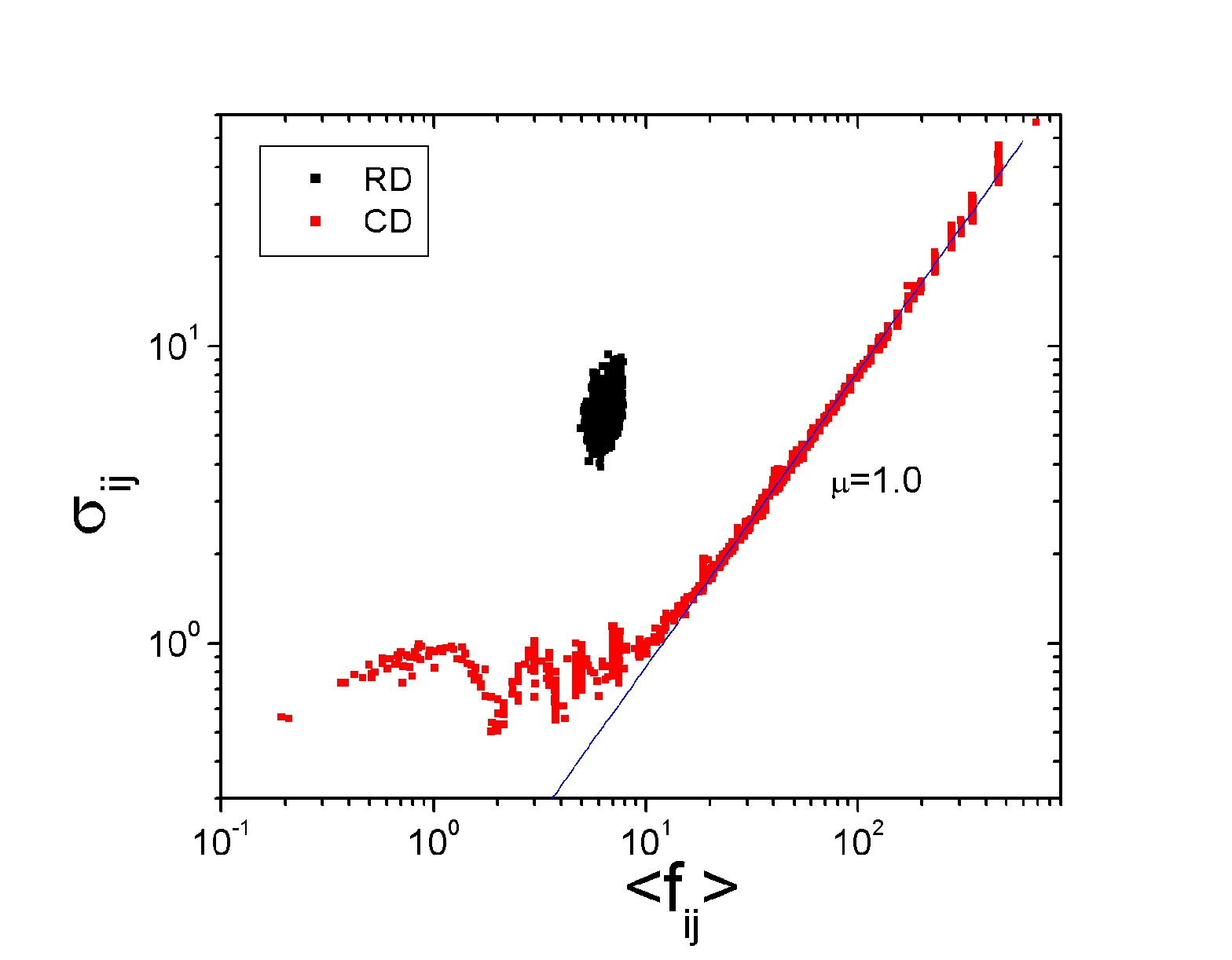}&
\includegraphics[width=7.0cm]{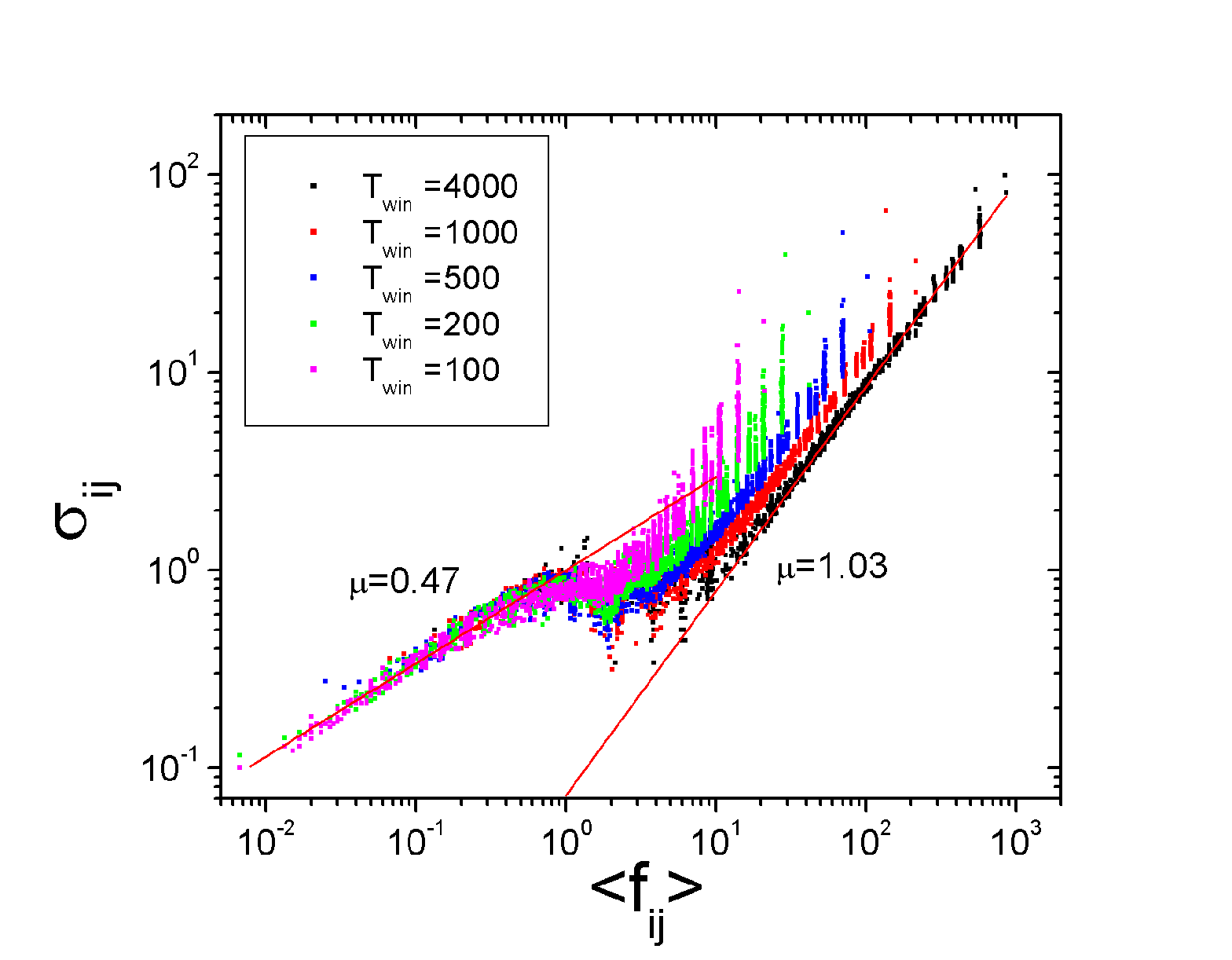}\\
{$(a)$} &{$(b)$}\\
\end{tabular}
\end{center}
\caption{(a) Dispersion $\sigma_{ij}$ against average $<f_{ij}>$ of
time series recorded at edges of the network within a fixed time
window $T_{WIN}=4000$ steps for random diffusion RD and CD
navigation rules. (b) Same as (a) but for different time windows and
CD navigation rule.} \label{fig-sigh-Links}
\end{figure}

\begin{figure}
\begin{center}
\includegraphics[width=6.5cm]{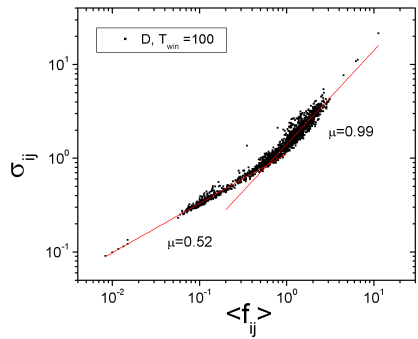} \caption{
The flow fluctuations $\sigma  _{ij}$ against average flow $<f_{ij}>$
for navigated diffusion with probabilistic preference for edges pointing to
low-connectivity nodes.} \label{Links-D}
\end{center}
\end{figure}

We apply a similar multi-channel analysis to the data on traffic flow
fluctuations. The flow along an edge between nodes $i$ and $j$ is
the sum of flow from $i\to j$ and from $j\to i$.  In Fig.\
\ref{fig-sigh-Links} we plot the standard deviations of the flow
time series $\sigma _{ij}$ against the average flow along that
edge $<f_{ij}>$. Each point on the plot represents one edge on the
network. As the Fig.\ \ref{fig-sigh-Links}a shows, in the case of CD
navigated diffusion, flow fluctuations obtained at a large number
of edges follow the same scaling pattern as that described in Eq.\
(\ref{eq-sigmah}) for noise fluctuations at nodes. However, in the
case of  edges, the situation becomes different when random
diffusion is used: the edges form a dense group on
the plot, representing almost equal fluctuation properties. This implies
that the dynamical preference in edges, which is built into the CD
navigation rules, introduces both an uneven  flow along
different links and a different fluctuation
pattern of the flow. The exponent that is measured from the data
in Fig.\ \ref{fig-sigh-Links}a is close to $\mu =1$. In the same
plot we find another group of edges for which the scaling exponent
cannot be defined. When the width
of the time window is reduced the fraction of
edges that belongs to the scaling regime with $\mu \approx 1$ is
reduced, and a number of edges appear in a new group,
for which the scaling exponent is close to $\mu \approx 0.5$.
The transition between these two groups is quite sharp for the
deterministic CD  navigation
 (see Fig.\ \ref{fig-sigh-Links}b). We investigate this question further by
considering a simplified, probabilistic
version of the edge-preferred navigation, called D algorithm \cite{ktr-libaptcn-06}, in which
a packet moves from node $i$ to its neighbour $j$ with probability
\begin{equation}
p_{ij}=1-\frac{k_j}{\displaystyle\sum_{j=1}^{N}C_{ij}k_j}
\label{d-func}
\end{equation}
where $C_{ij}$ is the adjacency matrix and as before $k_j$ is the degree of node $j$. Hence packets are more likely to move towards neighbours with a low connectivity.
With this navigation rule, we find the flow fluctuations, shown in Fig.\
\ref{Links-D}, exhibiting  two distinct groups of edges with
a smooth transition.

Study of the flow fluctuations suggests that the dynamical
preference of edges, as in our navigation rules, introduces the
necessary distinction between edges that leads to the scaling
behaviour. In contrast to the noise fluctuations studied above, in
flow fluctuations we did not observe a unique scaling exponent
with a smooth dependence of the acquisition time window. According
to their flow fluctuations, all edges appear to be in one of the
two groups of well defined scaling exponents, either 1/2 or 1. The
relative population of the groups changes, where the size of the
group with the
larger exponent is increasing with the length of the time window.

\section{Origin of Noise Scaling on Complex Networks}

In the previous section a comparative analysis of the fluctuations of noise and flow suggested that in the diffusion processes  a preference between network
elements, that can be achieved in an inhomogeneous network, is necessary for the scaling of the type in Eq.\ \ref{eq-sigmah} to occur.
In order to further clarify this point, we have analysed different geometries
including regular  and homogeneous substrates.

\begin{figure}[htb]
\begin{center}
\begin{tabular}{ccc}
\includegraphics[width=4.3cm]{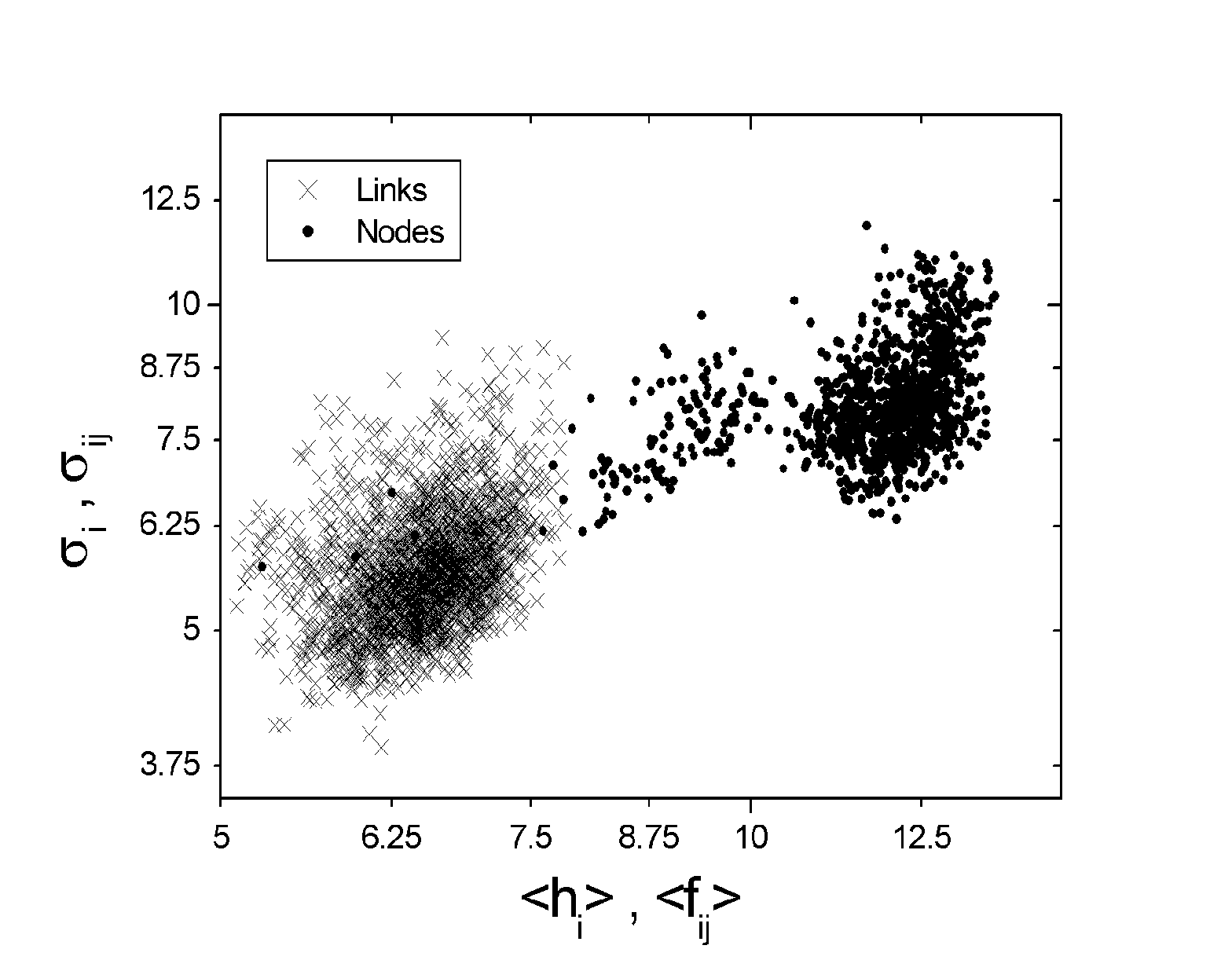} &
\includegraphics[width=4.3cm]{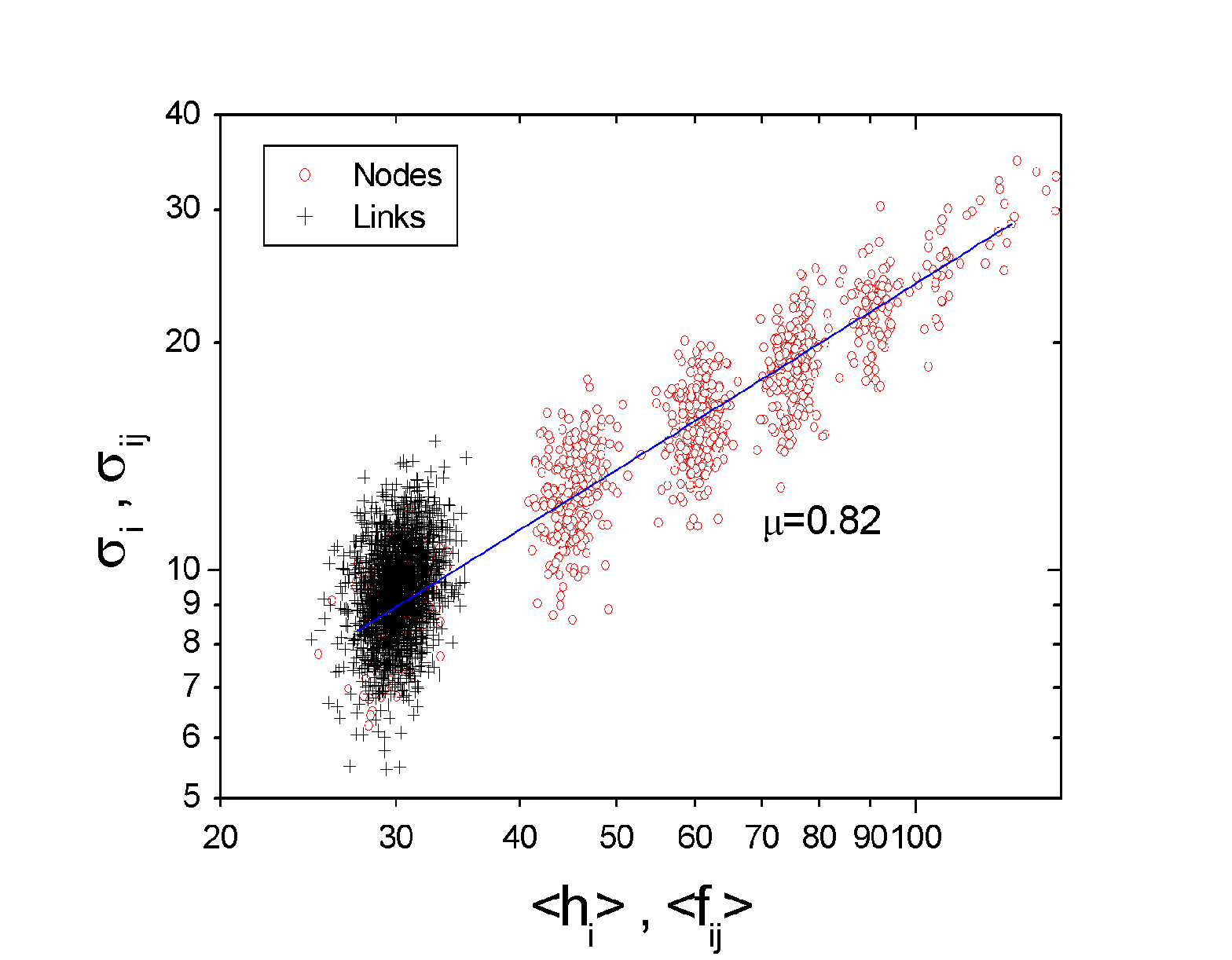} &
\includegraphics[width=4.3cm]{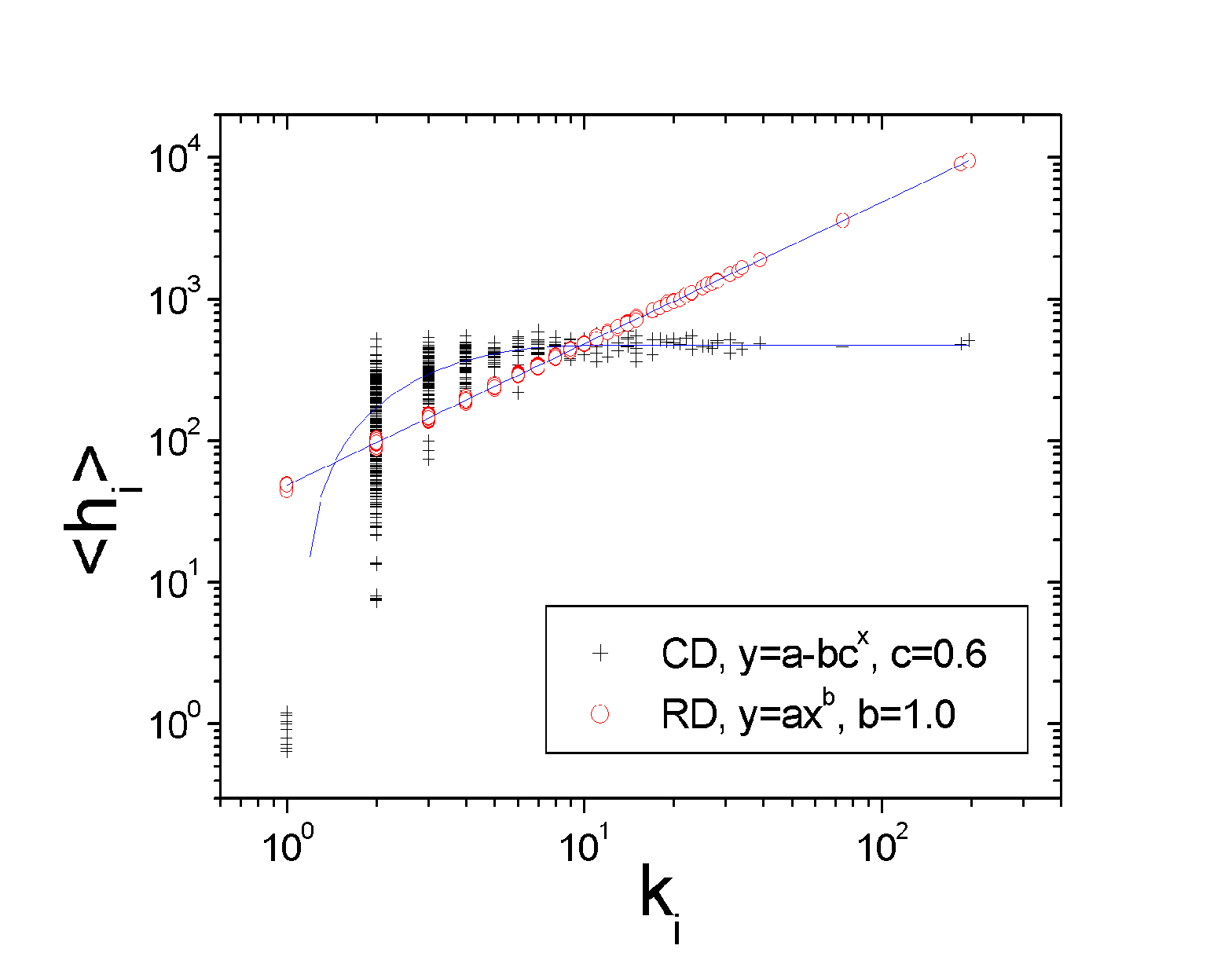} \\
{$(a)$} & {$(b)$} & {$(c)$}\\
\end{tabular}
\end{center}
\caption{For random diffusion, the
dispersion against average of time series recorded at
nodes and links on (a) regular square lattice and (b) random graph.
In both figures, the most left group of points represents the
data collected at (edges links), while the remaining groups are for nodes.
(c) The average number of packets processed by a node against node degree
for random  (RD) and  navigated (CD) diffusion.
}
\label{fig-sigh-regular-rg}
\end{figure}

In Fig.\ \ref{fig-sigh-regular-rg}a and b we show the fluctuations of packets in random diffusion recorded on nodes and edges of a square lattice and of a random graph.  Clearly, fluctuations of flow at edges on these structures appear to belong to one group within  a statistical dispersion.
However, the record of noise at different nodes already shows grouping according to their connectivity: on the
square lattice one can distinguish between fluctuations in the number of packets processed by  the four corner nodes, by nodes at the boundaries and
by the rest of the nodes, whose the connectivity is four. A larger range of groups of nodes is found on a random graph, with connectivities ranging
from 2 to 8 edges per node, where the larger connectivity
nodes are shifted to the right part of the plot.
In the same spirit, on the scale-free network the span of different groups of nodes, with different connectivities is even larger, also visible in Fig.\
\ref{fig-sigh-nodes}a, and it is related to the connectivity profile.
Therefore, the direct relationship between the node connectivity and the
number of processed packets by that node (its dynamic centrality) occurs
in the random diffusion. This represents  the basis of
the observed scaling of noise fluctuations on the scale-free networks.
In Fig.\ \ref{fig-sigh-regular-rg}c we show the
emergent linear dependence between the average noise at a node $<h_i>$ and its connectivity $k_i$ for our scale-free network,
which applies for the case of the random diffusion algorithm
(see also \cite{noh04}).
However, when the diffusion rules are changed, as in our navigated diffusion,
for example,
these relationships are altered. We find that in the case of CD navigation
the average number of packets at nodes of large connectivity plateaus
according to the functional form $<h_i> = a-bc^{k_i}$, with $c=0.6$.

Hence, nodes with different connectivity play a different role in the diffusion processes
on networks. This idea of dissimilarity can be extended to
the edges of the network when diffusion rules, such as our CD navigation, are implemented. Again, we observe distinct groups of edges which have different flow fluctuations,
with more 'important' groups having proportionally larger flow and large fluctuations.

It is an intrinsic  property of the diffusive  dynamics that different groups of nodes or edges develop a cooperative behavior and fall on a line with a well defined profile, as given in Eq.\ (\ref{eq-sigmah}).
Our results for the diffusive dynamics of information packets
suggest that stationarity of the time series is a necessary condition for such
cooperativity to occur, whereas the
dynamic continuity need not  be strictly observed (in our model sources and sinks of packets occur, which are balanced in the average).
 Our results show that in topologically or dynamically inhomogeneous systems nodes or edges
may develop different levels of cooperative behavior that leads to non-universal scaling.
The non-universality is represented  by the fact that the scaling exponent depends on the dynamics, with either a continuous change between the two limiting values
$0.5 \leq \mu \leq 1$, or with two exponents defined for different groups of
edges. A possible origin of these two limiting values of the scaling
exponent has  been discussed in
\cite{mb-fnd-04,da-otusftcn-06,ek-sttcsdftvs-06}.

\section{Return-Times to Nodes and Edges}

Another type of dynamic measure collected at individual nodes and
edges, that depends on the dynamic behaviour of the whole network,
is the statistics of return-times, or time intervals between the
successive events at a given node or an edge. In collective
dynamical systems such as earthquakes
\cite{c-ltcsutoeq-04,c-mrdbieqrt-05,c-ueqojctad-06}, critical
sandpiles \cite{bc-ltttlfsocgs-97,snc-wtssocs-02}, and stock market dynamics
\cite{skdr-wtdfm-02,llr-wtdsmi-06}, a broad distribution of return times
(sometimes called waiting times or recurrent times) is always
found, with power-law tails suggesting the occurrence of
long-range dynamic correlations between the events. In this work
we address the question of return times to nodes and to edges in
order to investigate further the nature of collective dynamic
behaviour in our model of diffusion of
packets on a scale-free network.

{\it Return-Time to Nodes.} In the case of random diffusion
(random walks) on networks the return time distribution has been studied in other parts of the theoretical physics literature. In
particular, the first return time to the origin of a random walker
on sparse random graphs, with nodes representing states of a
system, was considered by Bray and Rodgers \cite{br-dscsmgr-88} as a model of non-exponential relaxation in spin glasses and other non-ergodic
systems. With the help of some heuristic arguments, they arrived
at the conclusion that on a random graph the long-time behaviour in the diffusion in
the phase space is dominated by the parts of the network with
linear chains (no loops), leading  to the expression $P(\Delta t) \sim
\exp(-A(p)(\Delta t)^{1/3})$, where $p$ is the homogeneous
connectivity of the random graph and $A(p)$ is known.

\begin{figure}[htb]
\begin{center}
\begin{tabular}{cc}
\includegraphics[width=6.8cm]{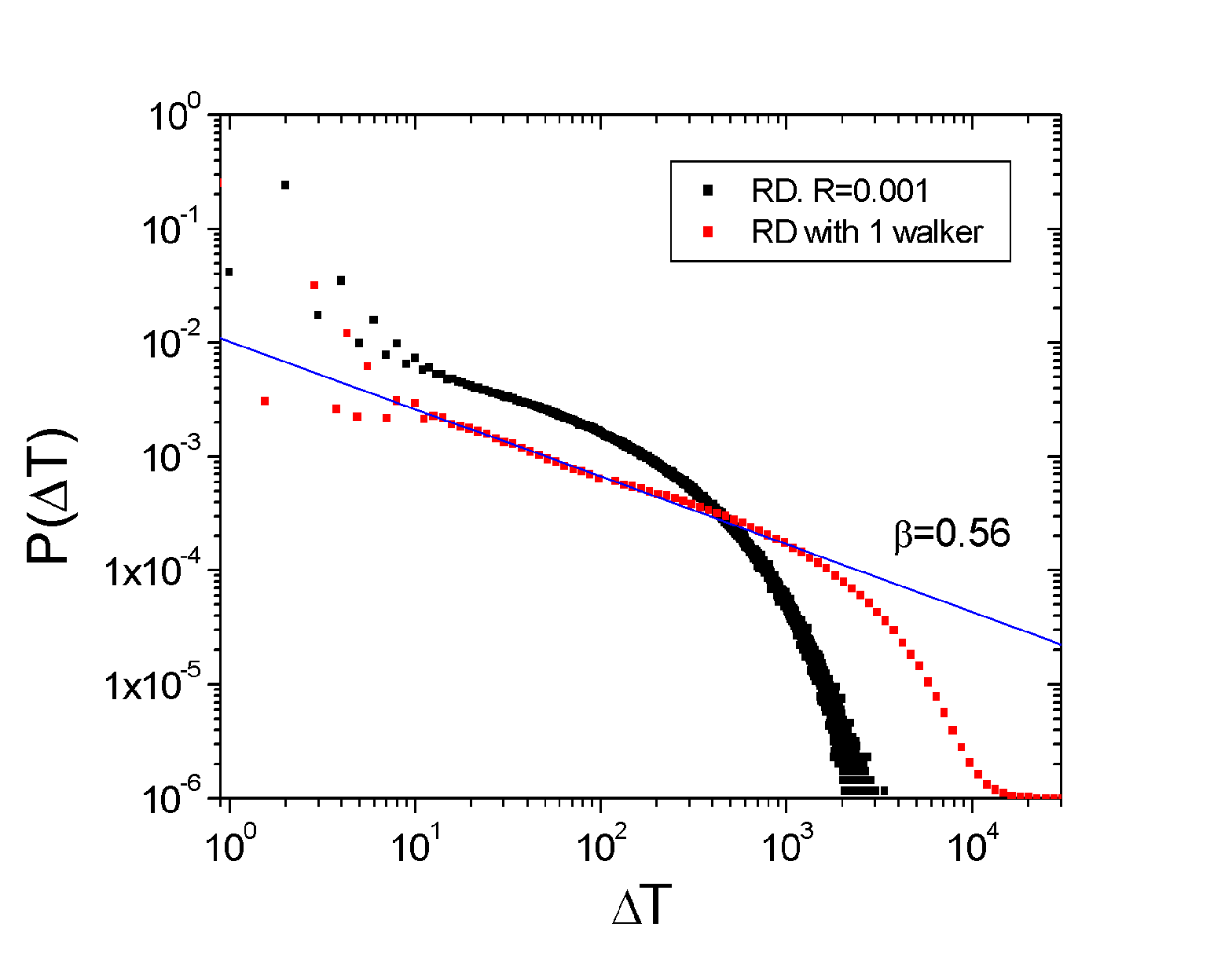}&
\includegraphics[width=6.8cm]{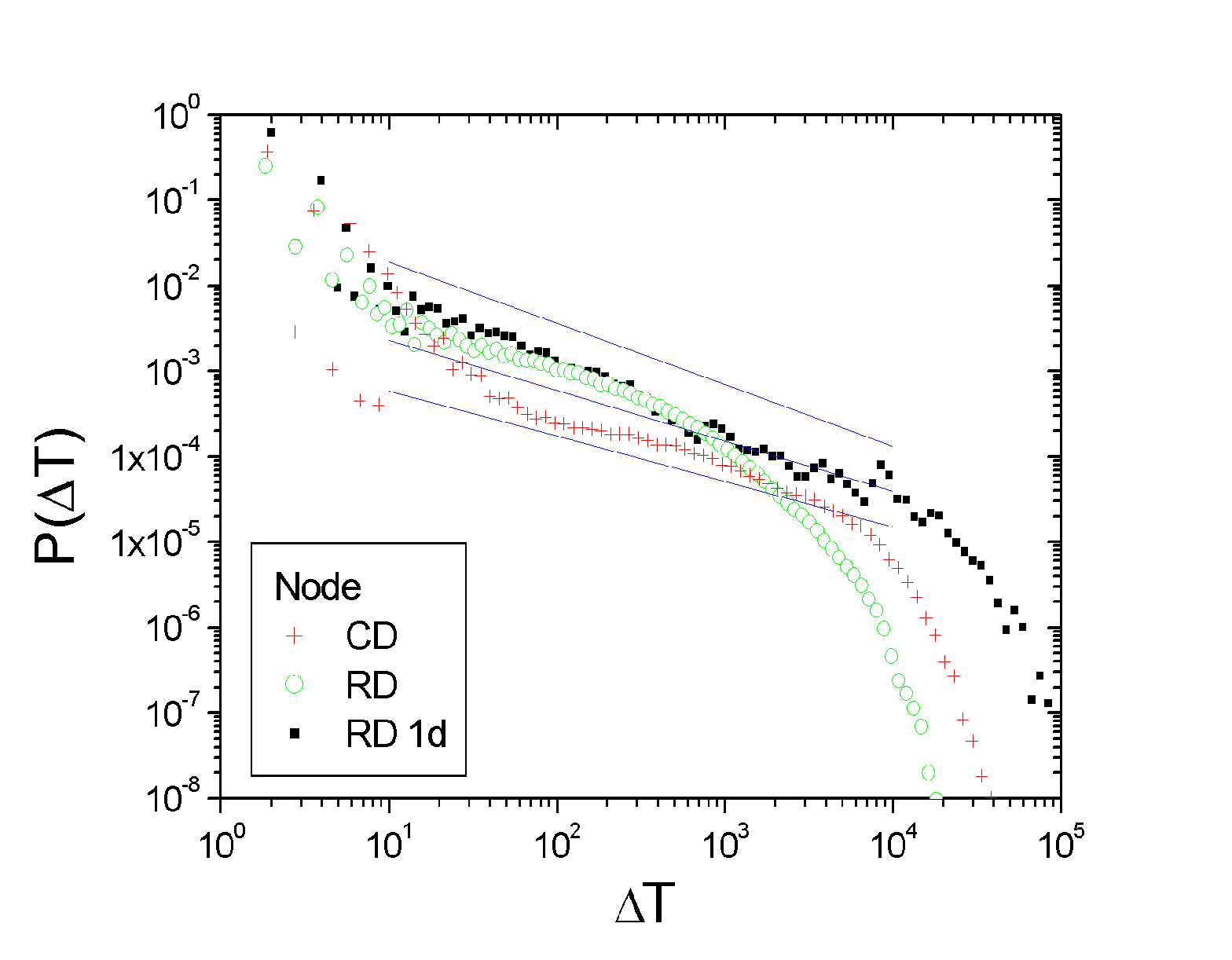}\\
{$(a)$} & {$(b)$} \\
\end{tabular}
\end{center}
\caption{The all-return time distribution to nodes in the
scale-free network for non-interacting and interacting walks with
(a) the random diffusion algorithm and (b) for interacting walks with
different navigation algorithms.} \label{fig-rettimes}
\end{figure}

These arguments can be generalized to introduce a power-law distribution of connectivities. If the distribution of $p$ behaves like $\sim p^{-\tau}$, and using the result in \cite{br-dscsmgr-88} that as $p \to 0$,
$P(\Delta t) \sim p \exp(-2\Delta t/p)$, then integrating over $p$  leads to $P(\Delta t) \sim (\Delta t)^{-\tau _{\Delta}}$ with
\begin{equation}
\tau_{\Delta} = \tau -2 \ .
 \label{tau-delta}
\end{equation}
Thus, the inhomogeneous connectivity creates a power-law
distribution in the return times distribution for RD. Recently a more rigorous treatment of random walks on scale-free networks
was carried out by Noh and Rieger \cite{noh04}, that yielded identical
results.
The results of our simulations for different diffusion processes
are shown in Fig.\ \ref{fig-rettimes}a. The return-time
distributions in different cases studied here seem to have a
power-law behaviour before a cut-off. (The cut-off can be
related to the network size in the case of single random walker.
Note also a characteristic splitting at small  $\Delta T$ with
 an inherent preference for even return times, caused by the
lack of clustering and the low density of walkers.)
In the case of non-interacting random walks, i.e., random diffusion
without queuing, the results agree, within error bars, with
the above theoretical prediction. We have the exponent
$\tau_{\Delta} = 0.56 \pm 0.03$, whereas the distribution of the
network's connectivity has a power-law exponent $\tau \approx 2.5$
(see Sec. 2).

Increasing the traffic density reduces the value of the cut-off,
but the slope remains practically unchanged.
However, when the navigated diffusion is considered, both the slope and
the cut-off of the distribution are changed. In Fig.\ \ref{fig-rettimes}b we
show the results for navigated diffusion with the CD algorithm,
random diffusion with
one-depth layer search and RD at low packet density.

\begin{figure}[htb]
\begin{center}
\begin{tabular}{cc}
\includegraphics[width=6.8cm]{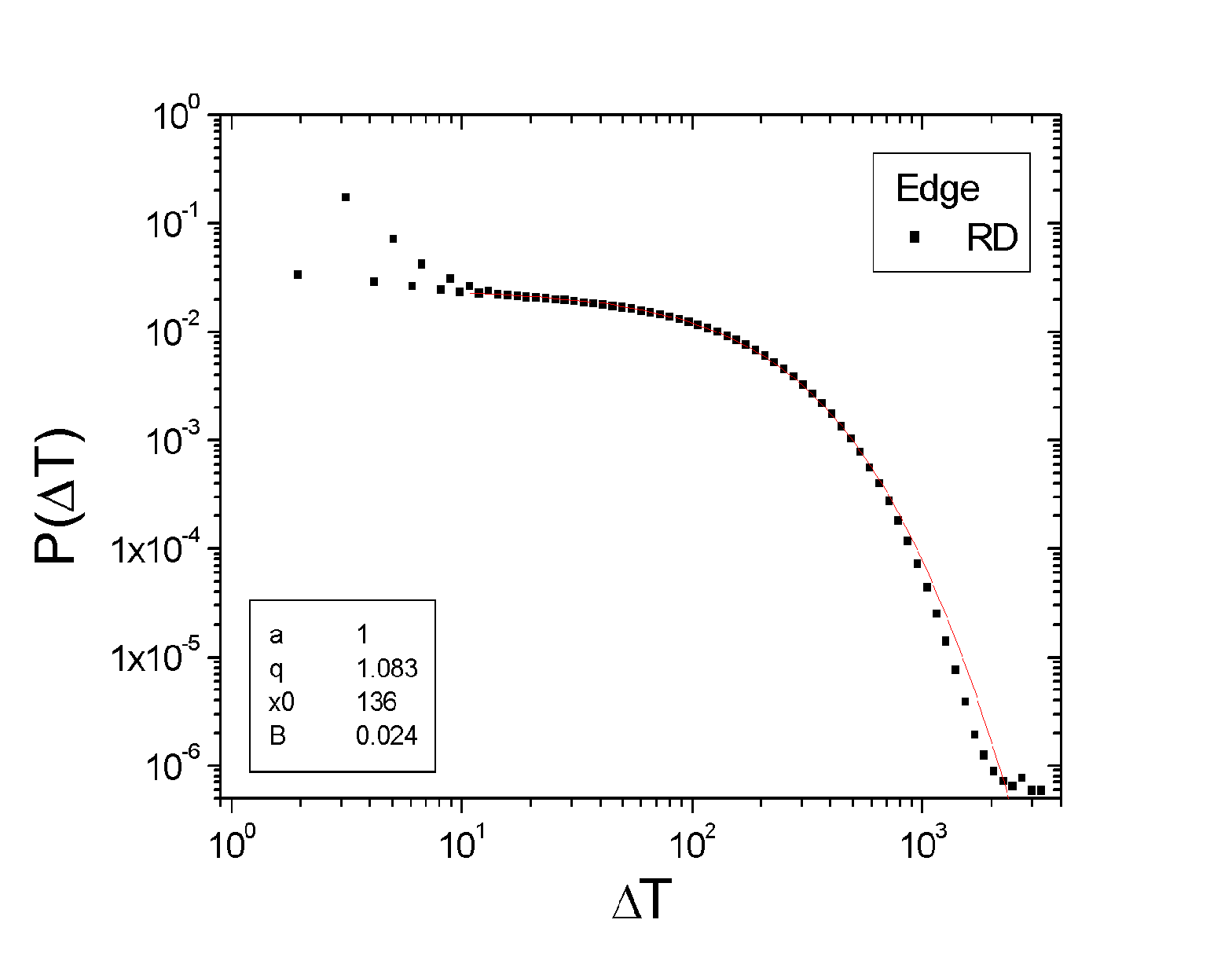}&
\includegraphics[width=6.8cm]{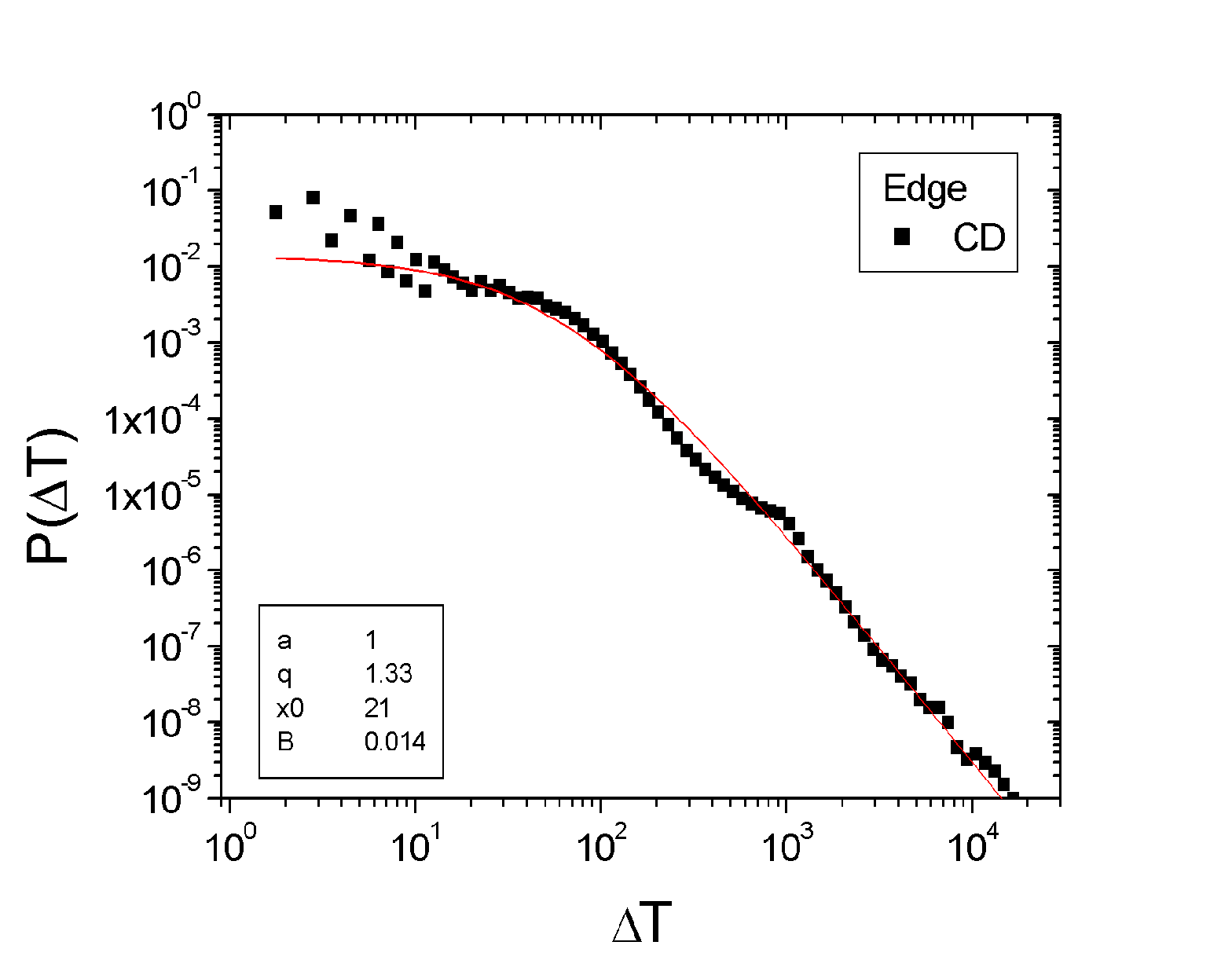}\\
{$(a)$} & {$(b)$} \\
\end{tabular}
\end{center}
\caption{The all-return time distribution to edges for (a) the random
diffusion and (b) CD-navigated diffusion algorithms. The fit lines
are explained in the text. } \label{fig-rettimes-edges}
\end{figure}

{\it Return-Time to Edges.} In contrast with the return times to nodes,
the situation is entirely different
from the point of view of edges on the same network. The results
are shown in Fig.\ \ref{fig-rettimes-edges}.  We find a pronounced
difference between the random and navigated diffusion
in the tails of the distributions. In both cases, however, a unique
functional form can be found.
 For larger $\Delta T$ the distributions of the return times to edges can be fitted  with a $q-$exponential form, which is often related to non-ergodic behaviour in dynamical systems \cite{tsallis88} :
\begin{equation}
P(\Delta t) = B\left(1 -(1-q){{\Delta t}\over{\Delta
t_0}}\right)^{1/(1-q)} \ .
\label{q-exp}
\end{equation}
In the case of random diffusion, shown in Fig.\ \ref{fig-rettimes-edges}a,
 the distribution
is very close to the exponential form, which corresponds to the $q\to 1$
limit of Eq.\ (\ref{q-exp}). In fact, we find $q=1.08$ in the case of
random diffusion, whereas
 in the case of edge-preferred CD navigation $q=1.33$, compatible with
 non-ergodic  dynamic behaviour.

\section{Conclusions and Discussion}

We have performed an analysis of the diffusive dynamics on a scale free network with
a power-law connectivity distribution but without any other form of
structural inhomogeneity.
Using the transport of information packets, where the diffusion
rules can be modified  in various
ways by adjusting the navigation of packets at nodes, we were able to
show that several dynamical effects appear to be related to the the
microscopic diffusion rules.
Our approach indicates that these findings will be relevant to more realistic transport problems on networks, which are
very different from simple random-walks.

In particular, we pinpointed the
importance of not just topological but also dynamic preference
to the occurrence of dynamic scaling.
We implemented navigation rules that involve preferences between
links, which is possible on topologically inhomogeneous scale-free
 networks.
The edge-preferential navigation rules appear to
dynamically homogenize the network (at large connectivity nodes)
and yield  new dynamical phenomena.
We focused on two types of scaling behaviour that can be obtained
from the point of view of individual structural elements, nodes
and edges of the network, and are potentially related to each other:
Scaling
of noise and flow fluctuations, on one hand, and scaling of
return-time distributions, on the other. While the noise fluctuations and/or
 return times at nodes have been studied extensively on different types of
networks \cite{mb-fnd-04,da-otusftcn-06,t-sfnfsfn-06,trt-tcnfjo-06}, our work
presents the first systematic study of the fluctuations
of flow on edges and return times to edges, and a comparison with the quantities
obtained at nodes within the same dynamics. Owing to the different roles that
nodes and edges play in these diffusion processes on an
inhomogeneous network, these comparative studies
lead to the conclusion that
certain types of preferential behaviour in either nodes or edges is
necessary for
the occurrence of  scaling. The scaling is characterized by the
exponents $\mu$  and $\tau_{\Delta}$ and the parameter $q$.

In particular, no scaling behavior was found in the
fluctuations of flow and in the return times to edges for random
diffusion.
On the other hand, when edge-preferred navigation is turned on, both
a non-trivial scaling of the flow fluctuations and a power-law tail in the
return-times to edges is observed.
The differences in node connectivities, studied in
regular, random graph, and the scale-free structures, leads to
differences in
the noise fluctuations both in random diffusion and in CD
navigated diffusion. Accordingly, we find non-trivial distributions of
the return times to nodes, that are power-law (up to a finite-size cut-off), in agreement with theoretical predictions \cite{br-dscsmgr-88,noh04}.
The exponents, however, depend
on the navigation rules. Only in the case of random diffusion do the
return times scale with a power which is  determined by
 the network's connectivity distribution.

Furthermore, we show that the scaling properties of noise fluctuations are
non-universal, with both
the topological inhomogeneity but also details of the
dynamics  playing a role in the emergent scaling
behaviour.  In addition to a well pronounced dependence on the width of the time window, that was also observed in other studies
\cite{da-otusftcn-06,t-sfnfsfn-06,trt-tcnfjo-06,ek-sttcsdftvs-06},
we found that
navigation rules effect the universality of noise fluctuations.
Again, we found a qualitative difference between the scaling properties at nodes and at edges. A unique  scaling exponent
$\mu (T_{WIN})$ is always found in case of noise fluctuations at nodes,
whereas in our
navigated diffusion the flow fluctuations at edges are bi-universal.

Finally, we have demonstrated that  by looking beyond the random walk
dynamics on sparse topologies, one may
find a number of new dynamical phenomena, that are both
interesting from the point of theory, but also practically important
for many real transport processes on networks.

{\bf Acknowledgments:} We acknowledge support from Programme
P1-0044 of the Ministry of Higher Education, Science and
Technology of the Republic of Slovenia, the British Council Partnerships in Science Project 22/2006 and the EC Marie Curie
Early Stage Training Programme NET-ACE (MEST-CT-2004-6724).

\bibliography{transport_main2}
\end{document}